\documentstyle[11pt,axodraw]{article}
\title{Analytical Structure Matching and\\
Very Precise Approach to the Coulombic\\
Quantum Three-Body Problem}

\author{TAN, Shi-Na \thanks{E-mail: tansn@itp.ac.cn}\\
{\it\small Institute of Theoretical Physics, CAS, P.O.Box 2735,
Beijing 100080, P.R.China}}

\date{}

\begin{document}

\maketitle

\begin{abstract}
A powerful approach to solve the Coulombic quantum three-body problem
is proposed. The approach is exponentially convergent and more efficient
than the Hyperspherical Coordinate(HC) method and the Correlation Function
Hyperspherical Harmonic(CFHH) method. This approach is numerically competitive
with the variational methods, such as that using the Hylleraas-type basis
functions. Numerical comparisons are made to demonstrate them, by calculating
the non-relativistic \& infinite-nuclear-mass
limit of the ground state energy of
the helium atom. The exponentially convergency of this approach is due to the
full matching between the analytical structure of the basis functions that
I use and the true wave function. This full matching was not reached by almost
any other methods. For example, the variational method using the Hylleraas-type
basis does not reflects the logarithmic singularity of the true wave function
at the origin as predicted by Bartlett and Fock. Two important approaches
are proposed in this work to reach this full matching: the coordinate
transformation method and the asymptotic series method. Besides these, this
work makes use of the least square method to substitute complicated numerical
integrations in solving the Schr\"{o}dinger equation, without much loss of
accuracy; this method is routinely used by people to fit a theoretical curve
with discrete experimental data, but I use it here to simplify the computation.

\vspace{3mm}
{\noindent
PACS number(s):}

\end{abstract}

\vspace{5mm}

\section{INTRODUCTION}

Most approximate methods to solve a linear partial differential
equation, such as the stationary state Schr\"{o}dinger equation,
are actually to choose an
$N$-dimensional subspace of the infinite-dimensional Hilbert space and
then to reduce the partial differential equation to $N$ linear algebraic
equations defined in this subspace. The efficiency of this kind of methods
is mainly determined by whether one can use sufficient small $N$ to reach
sufficient high accuracy, i.e., make the vector most close
to the true solution in this subspace sufficiently close to the true
solution while keeping the dimension $N$ not too large to handle.

Most methods to solve the Coulombic quantum three-body problem belong to
this class, except for some variational methods that make use of some
non-linear
variational parameters. The differences between different methods of this kind
mainly lie in different choices of the subspaces of the Hilbert space,
i.e., different choices of
the basis functions to expand the wave function.

Theoretically, any discrete
and complete set of basis functions may be used to expand the wave function,
and the convergency is easy to fulfilled. But actually, the convergency is
often slow and makes sufficient accuracy difficult to achieve. The naive
hyperspherical harmonic function method[1-3] in solving the Coulombic quantum
three-body problem is such an example--this slow convergency can be illustrated
by an analogous and simple example: to expand the function $f(x)=\sqrt{1-x^2}$
($-1\le x\le +1$) as a series of the Legendre polynomials of $x$. This series
is convergent like $N^{-s}$, where $s$ is a positive constant not large and $N$ is
the number of Legendre polynomials involved. The reason for this slow
convergency is that $f(x)$ is singular at $x=\pm 1$ but the Legendre
polynomials of $x$ are not. I call this the mismatching between the analytical
structures of the basis functions (the polynomials of $x$) and $f(x)$.

The correlation function hyperspherical harmonic(CFHH) method[4] were proposed to
overcome this difficulty. The spirit of this method can be simply illustrated,
still using the above example: to divide $f(x)$ by an appropriately selected
function(called the correlation function) to cancel the low order singularities
of $f(x)$ at $x=\pm 1$, then to expand the remaining function by the
Legendre polynomials of $x$. This time, the series is still convergent as
$N^{-s}$, but $s$ is increased by an amount depending on how many orders'
singularities have been canceled.

From this simple discussion one can see that the singularities of the function
$f(x)$ are not completely canceled by the correlation function, although more
sophisticated correlation function can cancel more orders' singularities.

A very simple approach to \emph{totally} eliminate the singularity is
to make an appropriate coordinate transformation,
and in the same time thoroughly give up the original hyperspherical
harmonic function method, not just repair it.
For example, for $f(x)=\sqrt{1-x^2}$, one may write $x=\sin{\theta}$, where
$-\pi/2\le\theta\le\pi/2$, then $f(x)=\cos{\theta}$ and one can expand $f(x)$
as the series about the Legendre polynomials of $(2/\pi)\theta$.
This time the series is \emph{factorially} convergent. The reason is that
the analytical structures of $f(x)$ and $P_l((2/\pi)\theta)$ match--they are
both analytical functions on the whole complex plane of $\theta$.

Another useful approach to solve this problem is to use the asymptotic series.
Still considering the example $f(x)=\sqrt{1-x^2}$, one may write the Taylor
series
$$f(x)=f_0+f_{1}x+f_{2}x^{2}+f_{3}x^{3}+\cdots~.$$
Of course, this series is slowly convergent near $x=\pm 1$. But one can use the
following asymptotic series to calculate $f_n$ when $n$ is large:
$$f_n=((-1)^n+1)(c_{3/2}n^{-3/2}+c_{5/2}n^{-5/2}+c_{7/2}n^{-7/2}+\cdots)~,$$
or, equivalently,
$$f_n=((-1)^n+1)\sum_{s=1/2}^{1/2+L}\tilde{f}_{s}\frac{s!}{n!(s-n)!}~,$$
where $s=\frac{1}{2}, \frac{3}{2}, \frac{5}{2}, \cdots, \frac{1}{2}+L$,
and $s!\equiv\Gamma(s+1)$. For a given $n \gg 1$, the error of
this formula is minimized
when $L/n \simeq 2/3$, and the minimum error is about 
$\sqrt{\frac{27}{2\pi^2}} n^{-2} 3^{-n}$, which exponentially decreases with $n$
increasing. Using such kind of asymptotic formulae to calculate the high order
coefficients of the Taylor series, one can expand the singular function $f(x)$
at high precision, with only finite linear parameters, $f_0, \cdots, f_n$ and
$\tilde{f}_{1/2}, \cdots, \tilde{f}_{1/2+L}$.

Now I introduce an alternative approach to reduce a differential equation to
a given finite dimensional subspace $\mathcal{L}_N$ of tbe Hilbert space.
Here $N$ is the dimension of the subspace. The central problem is how
to reduce an operator $O$ in the Hilbert
space, e.g., the kinetic energy operator or the potential energy operator,
to an $N\times N$ matrix in the given subspace. For a state
$\Psi\in\mathcal{L}_N$, the state $\Psi_O\equiv O\Psi$ usually
$\notin\mathcal{L}_N$. To reduce $O$ into an $N\times N$ matrix means to
find a state $\Xi\in\mathcal{L}_N$ to approximate $\Psi_O$. The usual approach
to select $\Xi$ is to minimize
$$(\Xi-\Psi_O, \Xi-\Psi_O)~,$$
where $(,)$ is the
innerproduct of the Hilbert space. This approach will reduce $O$ to a matrix
with elements
$$O_{ij}=(\phi_i, O\phi_j)~,$$
where $\phi_i\in\mathcal{L}_N$ is a set of orthonormal basis in $\mathcal{L}_N$,
satisfying $(\phi_i, \phi_j)=\delta_{ij}$, $1\le i, j\le N$. In numerical
calculation, the innerproduct is usually computed by numerical integration,
which needs sufficient accuracy and might be complicated. An alternative approach
that does not need these integrations is to write the states as
wavefunctions under a particular representation(e.g., the space-coordinate
representation), and then select $\Xi$ to minimize
$$\sum_{a}{|\Xi(x_{a})-\Psi_{O}(x_{a})|}^{2}~,$$
where $x_a$ is some sample points in the defining area of the wavefunctions.
In order to ensure $\Xi$ to be a good approximation of $\Psi_O$, the sample
points should be appropriately chosen. Usually the number of the sample points
is greater than and approximately proportional to $N$, and the separation between
two neighboring sample points should be less than the least quasi-semiwavelength
of a wavefunction in $\mathcal{L}_N$.

This alternative approach (I call it the least square method)
leads to a reduction of the operator $O$:
$$\tilde{O}_{ij}=(\tilde{\phi}_{i}, O\tilde{\phi}_{j})'~,$$
where $(,)'$ is a pseudo-innerproduct defined as
$(\phi,\psi)'\equiv\sum_{a}\phi^{*}(x_{a})\psi(x_{a})$ for arbitrary $\phi$
and $\psi$, and $\tilde{\phi}_i$ is a set of pseudo-orthonormal basis in
$\mathcal{L}_N$ satisfying $(\tilde{\phi}_{i}, \tilde{\phi}_{j})'=\delta_{ij}$.
We find that this approach is very similar to the usual one, except that
a discrete sum over sample points takes the place of the usual innerproduct
integration. And there is a great degree of freedom in the selection of the sample
points. In fact, as soon as the sample points are selected according to the
spirit mentioned above, the accuracy of the solution of the differential
equation usually will not decrease significantly. The major factor that determines
the accuracy of the solution is the choice of the subspace $\mathcal{L}_N$, which
has been discussed to some extent in previous pages.

In this work, solving the simpliest quantum three-body problem, the three methods
discussed above are all used: the coordinate transformation method,
the asymptotic series method, and the least square method. A high
precision is reached for the ground state energy of the ideal helium atom, and
the solution has also some merit in comparison with the Hyleraas-type variational
solution[5,6]. In section 2 the Bartlett-Fock expansion[7,8,9]
is studied, in order to
reflect the analytical structure of the wavefunction near the origin.
In this study, the asymptotic series are used to represent the hyper-angular
dependence of the wavefunction. In section 3 the $(u,w)$ coordinate system is
used to study the hyper-angular dependence of the wavefunction. This coordinate
system cancels the singularity of the hyper-angular functions totally. The
relationship between this coordinate system and the Hyleraas-type variational
method is also discussed. The least square method is used to 
reduce the hyper-angular parts of the kinetic
energy operator and the potential energy operator to finite-dimensional
matrices. In section 4 the connection
of the outer region solution and the inner region
Bartlett-Fock expansion is studied,
using the least square method. In section 5 the numerical result is presented
and compared with those of other methods. Some explanations are made. In section
6 some discussions are presented and some future developments are pointed out.

\section{BARTLETT-FOCK EXPANSION}

Considering an S state of an ideal helium atom, that is, assuming an infinite
massive nucleus and infinite light speed, one may write the Schr\"{o}dinger
equation
$$-2 t (\partial_{x}^2+\partial_{y}^2+\partial_{z}^2+\frac{1}{z}\partial_{z}) \psi
+ V \psi=E \psi~, \eqno (1)$$
where $x=r_{1}^{2}-r_{2}^{2}$, $y=2 r_{1} r_{2} \cos\theta_{12}$,
$z=2 r_{1} r_{2} \sin\theta_{12}$, and
$t = r_{1}^{2}+r_{2}^{2} = \sqrt{x^{2}+y^{2}+z^{2}}$. $r_1$ and $r_2$
are the distances of the electrons from the nucleus, and $\theta_{12}$
is the angle formed by the two electronic position vectors measured from the
nucleus. In this equation, an S state is assumed, so the wavefunction
$\psi$ is only dependent on $r_1$, $r_2$ and $\theta_{12}$, or, equivalently,
$x$, $y$, and $z$. The atomic unit, i.e.,
$\hbar=m_e=e^{2}/(4\pi\varepsilon_{0})=1$, is assumed throughout this paper.
The potential energy is
$$V=-\frac{2}{r_1}-\frac{2}{r_2}+\frac{1}{r_{12}}~, \eqno (2)$$
where $r_{12}$ is the distance between the two electrons.
$$r_{1}=\sqrt{\frac{t+x}{2}}~,~
  r_{2}=\sqrt{\frac{t-x}{2}}~,~
  r_{12}=\sqrt{t-y}~. \eqno (3)$$

\vspace{3mm}
The Bartlett-Fock expansion is
$$\psi=\sum_{n,k}\psi_{n,k}t^{n}\frac{(\ln t)^k}{k!}~, \eqno (4)$$
where $n=0, 1/2, 1, 3/2, 2, \cdots$, and $k=0, 1, 2, \cdots$.
$\psi_{n,k}$ only depends on the two hyper-angles, say,
$\alpha\equiv x/t$ and $\beta\equiv y/t$, and does not depend
on the hyper-radius, $\rho\equiv\sqrt{t}$. When $k>n$, 
$\psi_{n,k}\equiv 0$.

Using the coordinates $t$, $\alpha$, and $\beta$, one may rewrite
the Schr\"{o}dinger equation (1) as
$$(\partial_{t}^{2}+\frac{3}{t}\partial_{t}+\frac{1}{t^2}L_{0})\psi
=(v t^{-3/2} + p t^{-1})\psi~, \eqno (5)$$
where $p\equiv -E/2$, and
$$L_{0}=(1-\alpha^{2})\partial_{\alpha}^{2}
-2\alpha\beta\partial_{\alpha}\partial_{\beta}
+(1-\beta^{2})\partial_{\beta}^{2}
-3\alpha\partial_{\alpha}-3\beta\partial_{\beta}~; \eqno (6)$$
$$v=-\frac{\sqrt{2}}{\sqrt{1+\alpha}}-\frac{\sqrt{2}}{\sqrt{1-\alpha}}
+\frac{1/2}{\sqrt{1-\beta}}~. \eqno (7)$$

Substituting eq.(4) into eq.(5), and comparing the corresponding coefficients
before $t^{n}(\ln t)^k$, one will obtain
$$L_{n}\psi_{n,k}+(2 n+2)\psi_{n,k+1}+\psi_{n,k+2}
=v\psi_{n-\frac{1}{2},k}+p\psi_{n-1,k}~, \eqno (8)$$
where $L_{n}\equiv n(n+2)+L_{0}$.

The functions $\psi_{n,k}$ are solved out in the order
with n increasing; and for each n, with k decreasing. The physical
area of $(\alpha, \beta)$ is the unit circle: $\alpha^{2}+\beta^{2}\le 1$.
And the function $\psi_{n,k}(\alpha,\beta)$ may has singularities at
$\alpha=\pm 1$ and at $\beta=1$. The singularities are of these kinds:
$(1-\alpha)^{s}$, $(1+\alpha)^{s}$, and $(1-\beta)^{s}$, with
$s=\frac{1}{2}, \frac{3}{2}, \frac{5}{2}, \cdots$. So one may write
the Taylor series in the $(\alpha,\beta)$ unit circle:
$$\psi_{n,k}(\alpha,\beta)=\sum_{a,b=0}^{\infty}
\psi_{n,k,a,b}\alpha^{a}\beta^{b}~. \eqno (9)$$
The singularities make the usual cutoff, $a+b\le L_f+L_s$, inappropriate, because
the error decreases slowly when $L_f+L_s$ increases.
But since we have known the forms of the singularities, we can write the
asymptotic formulae to calculate those high order Taylor coefficients that
have important contributions:
$$\psi_{n,k,a,b}=\sum_{s=\frac{1}{2}}^{L_i-\frac{1}{2}}
\tilde{\psi}_{n,k,b,s}
{s \choose a}[1+(-1)^{a}]~; \eqno (10-1)$$
$$\psi_{n,k,a,b}=\sum_{s=\frac{1}{2}}^{L_i-\frac{1}{2}}
\tilde{\tilde{\psi}}_{n,k,a,s}
{s \choose b}(-1)^{b}~. \eqno (10-2)$$
Eq.(10-1) is appropriate when $a\gg b$ and $a\gg 1$,
while eq.(10-2) is appropriate when $b\gg a$ and $b\gg 1$.
${s \choose a}\equiv(s!)/[a!(s-a)!]$, and $s!\equiv\Gamma(s+1)$.
Here I have assumed the state is a spin-singlet, and thus
$\psi_{n,k}(-\alpha,\beta)=\psi_{n,k}(\alpha,\beta)$. For a spin-triplet,
the factor $[1+(-1)^{a}]$ in eq.(10-1) should be substituted by
$[1-(-1)^{a}]$.

In my actual calculation, the $(a,b)$ plane is divided into four areas:\\
the finite area: $0\le a,b\le L_f$ and $a+b\le L_{f}+L_{s}$ ($L_f\gg L_s\gg 1$),\\
the $a$-asymptotic area: $a>L_{f}$ and $b\le L_s$,\\
the $b$-asymptotic area: $b>L_{f}$ and $a\le L_s$,\\
and the cutoff area: the remain area.

Eq.(10-1) is used in the $a$-asymptotic area, and eq.(10-2) is used in the
$b$-asymptotic area, while the contribution from the cutoff area is neglected for
it is extremely tiny when $L_f\gg L_s\gg 1$.

In a word, a relevant hyper-angular function is described by a finite set of
parameters up to a high precision. These parameters are some Taylor
coefficients and some asymptotic coefficients.
To operate with some functions of this kind means
to operate with the corresponding sets of parameters. The relevant operations
are: addition of two functions--adding the corresponding parameters of the
two sets; multiplying a function by a constant--multiplying each parameter in the
set by the constant; multiplying a function by $v(\alpha,\beta)$(eq.(7))--
an appropriate linear transformation of the set of parameters of the multiplied
function; solving an equation $L_{n}f=g$ with g known and f unknown--solving
a set of linear equations about the parameters corresponding to $f$. Here, I write
the relevant linear equations corresponding to the equation $L_{n}f=g$:
$$[n(n+2)-(a+b)(a+b+2)]f_{a,b}+(a+1)(a+2)f_{a+2,b}+(b+1)(b+2)f_{a,b+2}
=g_{a,b}~; \eqno (11-0)$$
$$[n(n+2)-(b+s)(b+s+2)]\tilde{f}_{b,s}+(s+1)(2s+2b+3)\tilde{f}_{b,s+1}
+(b+1)(b+2)\tilde{f}_{b+2,s}=\tilde{g}_{b,s}~; \eqno (11-1)$$
$$[n(n+2)-(a+s)(a+s+2)]\tilde{\tilde{f}}_{a,s}
+(s+1)(2a+2s+3)\tilde{\tilde{f}}_{a,s+1}
+(a+1)(a+2)\tilde{\tilde{f}}_{a+2,s}
=\tilde{\tilde{g}}_{a,s}~. \eqno (11-2)$$

The detailed order to solve $\psi_{n,k}$ is:

Case 1: $n=[n]+\frac{1}{2}$, where $[n]$ is an integer. In this case,
solve $\psi_{n,[n]}$ from eq.($8_{n,[n]}$);
and then solve $\psi_{n,[n]-1}$ from eq.($8_{n,[n]-1}$);
$\cdots$; at last solve $\psi_{n,0}$ from eq.($8_{n,0}$). For each
$\psi_{n,k}$, the order is: first solve the asymptotic coefficients,
from $s=\frac{1}{2}$ to $s=L_{i}-{1 \over 2}$; then solve the
Taylor coefficients, from $a+b=L_{f}+L_{s}$ to $a+b=0$(i.e.,$a=b=0$).

Case 2: $n$ is an integer. In this case, the order is more complicated,
because the operator $L_{n}$ has zero eigenvalue(s) in this case. The order
is as following:

\vspace{0.5mm}
\noindent Step 1: \parbox[t]{146mm}
{set the asymptotic coefficients and the $a+b>n$ Taylor coefficients
of $\psi_{n,n}$ to zero;}

\vspace{1.5mm}
\noindent step 2: \parbox[t]{146mm}
{$k\gets n-1$;}

\vspace{1.5mm}
\noindent step 3: \parbox[t]{146mm}
{if $k<0$, goto step 8;}

\vspace{1.5mm}
\noindent step 4: \parbox[t]{146mm}
{solve the asymptotic coefficients and $a+b>n$ Taylor coefficients of
$\psi_{n,k}$, from eq.($8_{n,k}$), in the order analogous to that of case 1.}

\vspace{4mm}
\noindent step 5: \parbox[t]{146mm}
{solve the $a+b=n$ Taylor coefficients of $\psi_{n,k+1}$, from
eq.($8_{n,k}$).}

\vspace{1.5mm}
\noindent step 6: \parbox[t]{146mm}
{solve the $a+b<n$ Taylor coefficients of $\psi_{n,k+1}$, from
eq.($8_{n,k+1}$), with $a+b$ decreasing(analogous to case 1) to $0$.}

\vspace{4mm}
\noindent step 7: \parbox[t]{146mm}
{$k\gets k-1$, and goto step 3;}

\vspace{1.5mm}
\noindent step 8: \parbox[t]{146mm}
{set the $a+b=n$ Taylor coefficients of $\psi_{n,0}$ with some free
parameters;}

\vspace{1.7mm}
\noindent step 9: \parbox[t]{146mm}
{solve the $a+b<n$ Taylor coefficients of $\psi_{n,0}$,
from eq.($8_{n,0}$), with $a+b$ decreasing(analogous to case 1) to $0$.}

\vspace{4mm}
The free parameters in solving eq.(8)(see step 8 of case 2) are
finally determined by the boundary condition: $\psi\to 0$,
when $t\to +\infty$. In principle, we can use the Bartlett-Fock expansion
(eq.(4)) for arbitrary $t$, because it is always convergent. But actually,
when $t$ is large, the convergency is slow and there is canceling of large
numbers before this convergency is reached, both of which make
the Bartlett-Fock expansion impractical. So I only use this expansion when
$t$ is relatively small(see ref.[15] for similarity):
$\sqrt{t} \le \rho_{0}$.

In atual calculation,
I chose $L_{f}=100$, $L_{s}=20$, $L_{i}=6$, $n_{max}=7.5$ (the largest n value
of the terms in eq.(4) that are not neglected),
and $\rho_{0}=0.4$, and found that
the numerical error for the calculation
of the inner region ($\sqrt{t}\le\rho_{0}$)
wavefunction is no more than a few parts in $10^{10}$.
I use this method to test the accuracy of the calculation: set $E$ in
eq.(8) (note that $p \equiv -E/2$) equal to an initial value
(for example, set $E_{initial} = -2.9037$, or set $E_{initial} = -2.903724377$),
and use the approximate wavefunction $\psi_{app}$
thus obtained to calculate the value
$(H\psi_{app})/\psi_{app}$, where $H$ is the exact Hamiltonian operator,
and I find it to be almost equal to the initial value $E_{initial}$,
with a relative error no more than a few parts in $10^{10}$.

When $t$ is larger, another approach is used:

\section{THE HYPER-ANGULAR DEPENDENCE OF THE WAVEFUNCTION}

We have seen that the hyper-angular dependence of the wavefunction,
described as a function of $(\alpha,\beta)$ for each fixed
$\rho\equiv\sqrt{t}\equiv\sqrt{r_{1}^2+r_{2}^2}$, has
singularities at $\alpha=\pm 1$ and at $\beta=1$. Physically,
this corresponds to the case that the distance between two of the three
particles equals zero. It can be proved that,
for a spin-singlet, the following coordinate
transformation will eliminate these singularities \emph{thoroughly}:
$$u=\sqrt{\frac{1+\alpha}{2}}+\sqrt{\frac{1-\alpha}{2}}-1~,~
w=\sqrt{1-\beta}~. \eqno (12)$$
Equivalently,
$$u=\frac{r_{1}+r_{2}}{\rho}-1~,~w=\frac{r_{12}}{\rho}~. \eqno (13)$$
If the energy-eigenstate $\psi$ is symmetric under the exchange
of $r_1$ and $r_2$(spin-singlet),
I believe that, for each fixed $\rho$, $\psi$ is a \emph{entire}
function of $(u,w)$.
If the energy-eigenstate $\psi$ is antisymmetric under the interchange
of $r_1$ and $r_2$(spin-triplet), I believe that, for each fixed $\rho$,
$\psi=\frac{r_{1}-r_{2}}{\rho}\phi$, where $\phi$ is a \emph{entire}
function of $(u,w)$.

This beautiful characteristic makes it especially appropriate to approximate
$\psi$, for each fixed $\rho$, by an $n$-order polynomial of $(u,w)$, not by
an $n$-order polynomial of $(\alpha,\beta)$. The former expansion, a polynomial
of $(u,w)$, matches the analytical structure of $\psi$; while the latter one,
a polynomial of $(\alpha,\beta)$, does not. The hyper-spherical harmonic
function method belongs to the latter expansion, a polynomial of
$(\alpha,\beta)$. So the hyper-spherical harmonic function expansion does not
correctly reflect the analytical structure of $\psi$. The slow convergency
of the hyper-spherical harmonic function expansion is only a consequence of
this analytical structure mismatching.

We expect that the $(u,w)$ polynomial expansion converges \emph{factorially}
to the true wavefunction. It is worthful to demonstrate a similar example
to illustrate this. Consider a function $f(x)=\exp(-x),~-1\le x \le +1$;
expand $f(x)$ by Legendre polynomials:
$f(x)\doteq\sum_{l=0}^{n}f_{l}P_{l}(x)$; it can be proved that the error
of this formula is of the order $1/(2^{n}n!)$, which factorially approach zero
as $n$ increases.

Using the $(\rho,u,w)$ coordinates, one can write the Schr\"{o}dinger
equation as:
$$-{1\over 2}(\partial_{\rho}^2+{5\over\rho}\partial_{\rho}
+\frac{4L_0}{\rho^2})\psi+{C\over\rho}\psi=E\psi~, \eqno (14)$$
where $L_0$ and $C$ are the hyper-angular parts of the kinetic energy and
the potential energy, respectively.
$$4L_{0}=(1-2u-u^2)\partial_{u}^{2}+(2-w^{2})\partial_{w}^{2}
-\frac{2(1+u)(1-2u-u^{2})}{u(2+u)}\frac{(1-w^{2})}{w}\partial_{u}\partial_{w}
+\frac{(1+u)(4-10u-5u^{2})}{u(2+u)}\partial_{u}
+\frac{4-5w^{2}}{w}\partial_{w}~; \eqno (15)$$
$$C=-\frac{4(1+u)}{u(2+u)}+\frac{1}{w}~. \eqno (16)$$

The physical area $\mathcal{D}$ of $(u,w)$ is:
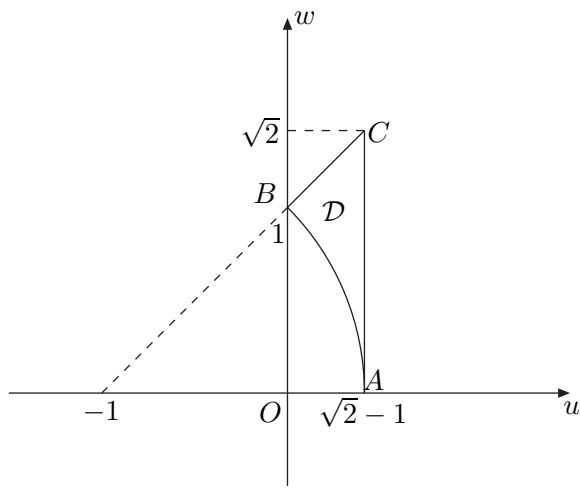
\vspace{5mm}


\begin{center}\begin{picture}(210,175)(0,0)

\LongArrow(0,35)(210,35)
\Text(213,32)[t]{$u$}
\LongArrow(105,0)(105,175)
\Text(108,177)[l]{$w$}
\Text(99,28)[]{$O$}
\Line(134,35)(134,134)
\Line(134,134)(105,105)
\DashLine(105,105)(35,35){3}
\DashLine(134,134)(105,134){3}
\CArc(35,35)(99,0,45)
\Text(35,28)[]{$-1$}
\Text(134,28)[]{$\sqrt{2}-1$}
\Text(102,99)[t]{$1$}
\Text(95,134)[]{$\sqrt{2}$}
\Text(123,104)[]{$\mathcal{D}$}
\Text(138,40)[]{$A$}
\Text(140,134)[]{$C$}
\Text(97,111)[]{$B$}

\end{picture}\end{center}

\vspace{1mm}
\noindent In this figure, point $A$ corresponds to the coincidence of the
two electrons, and point $B$ corresponds to the coincidence of the nucleus
and one electron.

For a spin-singlet,
we can use an n-order polynomial of $(u,w)$ to approximate $\psi$.
The coefficients of this polynomial are functions of $\rho$.
Denote by $\mathcal{L}_N$ the set of all the polynomials of $(u,w)$
with order no more than $n$. Here, $N=(n+1)(n+2)/2$ is the dimension.
In the physical area $\mathcal{D}$, I choose a set of points as sample points:
$$w_{a}=\sqrt{2} \frac{(a_{2}+0.5)}{n_{2}}~, \eqno (17)$$
$$u_{a}=(\sqrt{2}-1)-[(\sqrt{2}-1)-m(w_{a})] \frac{(a_{1}+0.5)}{n_{1}}~,
\eqno (18)$$
where $m(w)$ is the minimum physical $u$ value for a $w$ value.
$m(w)=\sqrt{2-w^{2}}-1$, if $w<1$; and $m(w)=w-1$, if $w\ge 1$.
$a\equiv (a_{1},a_{2})$, and $0\le a_{1}<n_{1}$, $0\le a_{2}<n_{2}$.
I chose $n_{1}=n_{2}=2n$, so there are altogether $4n^{2}$ sample points.
These sample points define a pseudo-innerproduct.
I constructed a set of
pseudo-orthonormal basis in $\mathcal{L}_N$,
by using the Schmidt orthogonalization method, and then reduce the operators
$L_0$ and $C$ to $N\times N$ matrices under this basis,
using the method introduced in section 1.

\section{CONNECTION OF THE INNER REGION AND THE OUTER REGION WAVEFUNCTIONS}

In the area $\rho<\rho_{0}$(inner region), the Bartlett-Fock expansion is used.
In the area $\rho>\rho_{0}/2$(outer region), $\psi$ is approximated
by a vector in $\mathcal{L}_N$ for each given $\rho$, and the partial
derivatives with respect to $\rho$ are substituted by optimized
variable-order and variable-step differences, which requires the selection
of a discrete set of $\rho$ values.
The overlap region of the inner region and the
outer region ensures the natural connection of the derivative of $\psi$,
as well as the connection of $\psi$ itself. The connection is performed
by using the least square method:
for a polynomial of $(u,w)$ at $\rho=\rho_{0}$, appropriately
choose the values of the free parameters of the solution of eq.(8)
(see section 2) so that the sum of the squares of the differences of the
the inner region solution and the outer region polynomial at the sample points
is minimized. This defines a linear transformation to calculate the values
of those free parameters from the given polynomial. When the values
of these free parameters are determined, one can calculate the values
of $\psi$ in the region $\rho_{0}/2<\rho<\rho_{0}$, using the Bartlett-Fock
solution, and further use these $\psi$ values to construct polynomials
of$(u,w)$ at $\rho_{0}/2<\rho<\rho_{0}$ (according to the law of
least square), and then use these polynomials
in the difference calculation of the partial derivative of $\psi$ with
respect to $\rho$ at $\rho\ge\rho_{0}$. At a sufficient large value
$\rho=\rho_{1}$, the first-class boundary condition is exerted; of course,
future development may substitute this by a connection with the long range
asymptotic solution of $\psi$.

At last, the whole Schr\"{o}dinger equation is reduced to an eigen-problem
of a finite-dimensional matrix. The dimension of the matrix is
$N_{\rho}\times N$, where $N_{\rho}$ is the number of free 
$\rho$ nodes
used in discretizing the partial derivatives with respect to $\rho$, and $N$
is the number of independent hyper-angular polynomials used. Note that
the energy value should be used in solving eq.(8), but it is unknown. The
actual calculation is thus an iteration process: choose an initial value
of $E_0$ to solve eq.(8) and form the $N_{\rho}\times N$ dimensional matrix,
and calculate the eigenvalue of this matrix to get a new value $E_1$, etc..
The final result is the fixed point of this iteration process. In actual
calculation, I found that the convergency of this iteration process
is very rapid if $\rho_0$ is relatively small. Choosing $\rho_{0}=0.4$,
I found that each step of iteration cause the difference between the eigenvalue
of the matrix and the fixed point decrease by about $(-160)$ times, when
calculating the ground state.

\section{NUMERICAL RESULT AND COMPARISONS}

Using 20 independent Bartlett-Fock series(up to the $t^{7.5}$ term in eq.(4),
neglecting higher order terms),
choosing $n=10$ (so that $N=66$), choosing $N_{\rho}=40$, with
$\rho_{0}=0.4$ and $\rho_{1}\doteq 11.32$, and
with the discrete values of $\rho$
equal to $0.4/1.2^{3}, 0.4/1.2^{2}, 0.4/1.2, 0.4, 0.4\times 1.2,
0.4\times 1.2^{2}, 0.4\times 1.2^{3}, \cdots, 0.4\times 1.2^{8}\doteq 1.7199,
0.4\times 1.2^{8}+0.3, 0.4\times 1.2^{8}+0.6, 0.4\times 1.2^{8}+0.9,
\cdots, 0.4\times 1.2^{8}+9.3$, and
$0.4\times 1.2^{8}+9.6\doteq 11.32$ (the first three
points are for the natural connection of the derivative of $\psi$,
the last point is for the first-class
boundary condition, and the remained 40 points are free nodes),
and discretizing the partial derivatives with respect to $\rho$ according
to the complex-plane-division rule(that is: when calculating the partial
derivatives with respect to $\rho$ at $\rho=l$, use and only use
those node points satisfying $\rho>l/2$ in the difference format, because the
point $\rho=0$ is the singular point), I obtained the
result for the ground state
energy of the ideal helium atom:
$$E=-2.9037243738~, \eqno (19)$$
compared with the accurate value:
$$E=-2.9037243770~. \eqno (20)$$
So the relative error of the result (19) is about $1.1\times 10^{-9}$.
Since my method is not a variational method,
the error of the approximate wavefunction
that I obtained should be of a similar order of magnitude, so if one
calculate the expectation value of the Hamiltonian under this approximate
wavefunction, the accuracy of the energy will be further raised by
several orders of magnitude.

The result (19) is much more accurate than the result
of ref.[10]:$-2.90359$, which
used the hyper-spherical coordinate method.
In ref.[10], the quantum numbers $(l1,l2)$
(angular momenta of the two electrons) are used and a cutoff for them
is made; this cutoff does not correctly reflect the analytical
structure of $\psi$ at $r_{12}=0$ (equivalently $\beta=1$). This is the major
reason causing the inaccuracy of the result of ref.[10].

It is also worthful to compare my result with that of ref.[4], in which
the correlation function hyper-spherical harmonic method is used. Note that
the result (19) is obtained by using a set of $N=66$
hyper-radius-dependent coefficients to expand the wavefunction. For a similar
size in ref.[4], N=64, the result is $-2.903724300$, with relative error
about $26.5\times 10^{-9}$. When N=169, the result of ref.[4] is
$-2.903724368$, with relative error about $3.1\times 10^{-9}$. Apparently
my method converges more rapidly than that of ref.[4]. The major reason
is that the correlation function hyper-spherical harmonic method does not
cancel the singularities totally---there is still some discontinuity
for the higher order derivatives, although the low order singularities,
which trouble the naive hyperspherical harmonic method, are canceled by
the correlation function.

\section{CONCLUSIONS, DISCUSSIONS AND FUTURE DEVELOPMENTS}

In conclusion, there are several important ideas in my work that
should be emphasized: first, I use the asymptotic series to compute
the Bartlett-Fock series up to a high precision, with error no more than,
for example, a few parts in $10^{10}$. Second, I propose an alternative
coordinate system, the $(u,w)$ system, in which the hyper-angular
singularities are thoroughly eliminated, which renders a factorial
convergency for the expansion of the hyper-angular function. Third,
I make use of the least square method to reduce an operator(infinite
dimensional matrix) to a finite dimensional matrix in a finite dimensional
subspace of the Hilbert space and to connect the solutions in different
regions, avoiding complicated numerical integrations, without much loss
of the accuracy for the solution. Fourth, the optimized difference format
---the complex plane division rule---is used to discretize the partial
derivatives of the wavefunction with respect to $\rho$. I calculated the
ground state energy of an ideal helium atom concretely and obtained a very
high precision, demonstrating that my method is superior to many other
methods and competitive with any sophisticated methods.

About the analytical structure of the stationary wavefunction:
1. there are logarithmic singularities at $\rho=0$, in the forms
of $\rho^{m}(\ln\rho)^{k}$; 2. for a given $\rho$, $\psi$ (for a spin-singlet)
or $\psi/[(r_{1}-r_{2})/\rho]$(for a spin-triplet) has no singularity,
as a function of $(u,w)$.

Here, I must mention the well known variational method based on
the Hyleraas-type functions, because it also
satisfies the second characteristic of
the wavefunction mentioned in the above paragraph.
One can see this by
a simple derivation. The Hyleraas-type function is a entire function
of $r_1$, $r_2$ and $r_{12}$, or equivalently, a entire function
of $r_{1}+r_{2}$, $r_{1}-r_{2}$, and $r_{12}$. For a fixed $\rho$,
one can substitute $(r_{1}-r_{2})^2$ in this function by
$2\rho^{2}-(r_{1}+r_{2})^{2}$, so that, for fixed $\rho$,
the function is a entire function
of $r_{1}+r_{2}$ and $r_{12}$ for spin-singlet, or such kind of entire
function times a common factor $r_{1}-r_{2}$ for spin-triplet. Equivalently,
for fixed $\rho$, the Hyleraas-type function is a entire function of
$(u,w)$(spin-singlet) or such kind of entire function times
$\frac{r_{1}-r_{2}}{\rho}$(spin-triplet).
This characteristic is one of the most important reasons that
account for the high accuracy of the Hyleraas-type variational method.

But this variational method also has its shortcoming: the Hyleraas-type
function does not reflect the logarithmic singularities with respect to
$\rho$. So, although this method has high precision for the energy levels,
the approximate wavefunctions that it renders may deviate significantly
from the true wavefunctions near the origin. See ref.[4,13,14] for detailed
discussions.

A central idea of this paper is: devising the calculation method
according to the analytical structure of the true solution. The $(u,w)$
coordinates, the Bartlett-Fock expansion and the asymptotic series approach
to compute this expansion, and the complex-plane-division rule in calculating
the partial derivatives with respect to $\rho$, all reflect this central
idea. The basic principle that ensures high numerical precision is just
this idea.

This preliminary work is incomplete in the following aspects:

First, how to \emph{prove} that $\psi$(for spin-singlet, or 
$\psi/[(r_{1}-r_{2})/\rho]$ for spin-triplet) has no singularity for fixed $\rho$,
as a function of $(u,w)$? Note that if this function still has singularities outside
of the physical area $\mathcal{D}$(see previous figure), the convergency of
the expansion of the hyper-angular function
will be only exponential, not factorial. Of course, even if
such kind of singularities do exist, my method will still converge more rapidly
than the correlation function hyperspherical harmonic method, because the latter
method only converges like $N^{-p}$, slower than $\exp(-\gamma\sqrt{N})$.
The rapid convergency of my method make me guess that such kind of singularities
do not exist.

Second, the asymptotic behavior of the wavefunction, when one electron is
far away from the nucleus, is not studied in this work. This problem
will be important when the highly excited states and the scattering states are
studied, a topic that will become my next object. 

Third, how to use the ideas proposed in this work to study a helium atom
with finite nuclear mass? Besides this, the relativistic and QED corrections
must be calculated, if one want to obtain a result comparable with
high-precision experiments.

Fourth, I have focused on the S states till now. When the total angular momentum
is not zero, there might be more than one distance-dependent functions (see, for
example, ref.[11]). I believe that some important analytical structures of the
S states studied in this work are also valid for those functions.

Surely, some important aspects of this work will also play an important
role in the highly excited states and the scattering states: the logarithmic
singularities about $\rho$ and the method to compute the Bartlett-Fock expansion,
the non-singularity with respect to the coordinates $(u,w)$, and the technique
to connect solutions of different regions, etc.. They can be applied
to the study of the highly excited states and the scattering states.

\vspace{5mm}
\noindent {\large\bf ACKNOWLEDGEMENTS}

\noindent The encouraging discussions with Prof. SUN Chang-Pu and with
Prof. Zhong-Qi MA are gratefully
acknowledged. I thank Prof. HOU Boyuan for providing me some useful references.
I am grateful to Prof.~C.M.~Lee~(Jia-Ming~Li) and Dr.~Jun~Yan for
their attention to this work and their advices.


\begin{thebibliography}{99}

\bibitem{1} V.B.Mandelzweig, Phys.Lett.{\bf A.78}, 25 (1980)

\bibitem{2} M.I.Haftel, V.B.Mandelzweig, Ann.Phys.{\bf 150}, 48 (1983)

\bibitem{3} R. Krivec, Few-Body Systems, {\bf 25}, 199 (1998)

\bibitem{4} M.I.Haftel, V.B.Mandelzweig, Ann.Phys.{\bf 189}, 29-52 (1989)

\bibitem{5} E.A.Hylleraas and J.Midtdal, Phys.Rev.{\bf 103}, 829 (1956)

\bibitem{6} K.Frankowski and C.L.Pekeris, Phys.Rev.{\bf 146}, 46 (1966)

\bibitem{7} J.H.Bartlett, Phys.Rev.{\bf 51}, 661 (1937)

\bibitem{8} V.A.Fock, Izv.Akad.Nauk SSSR, Ser.Fiz.{\bf 18}, 161 (1954)

\bibitem{9} J.D.Morgan, Theor.Chem.Acta{\bf 69}, 81 (1986)

\bibitem{10} Jian-zhi~Tang, Shinichi~Watanabe, and Michio~Matsuzawa,
Phys.Rev.{\bf A.46}, 2437 (1992)

\bibitem{11} W.T.Hsiang and W.Y.Hsiang, {\it On the reduction of the
Schr\"{o}dinger's equation of three-body problem to a system of linear
algebraic equations}, preprint (1998)

\bibitem{12} Zhong-Qi Ma and An-Ying Dai, {\it Quantum three-body problem},
preprint, physics /9905051 (1999); Zhong-Qi Ma, {\it Exact solution to the
Schr\"{o}dinger equation for the quantum rigid body},
preprint, physics /9911070 (1999).

\bibitem{13} J.H.Bartlett {\it et al.}, Phys.Rev.{\bf 47}, 679 (1935)

\bibitem{14} M.I.Haftel and V.B.Mandelzweig, Phys.Rev.{\bf A.38}, 5995 (1988)

\bibitem{15} James~M.Feagin, Joseph~Macek and Anthony~F.Starace,
Phys.Rev.{\bf A.32}, 3219 (1985)
\end{thebibliography}
\end{document}